
\tolerance=10000
\input phyzzx

\font\mybb=msbm10 at 12pt
\def\bbcc#1{\hbox{\mybb#1}}

\def\R {\bbcc{R}}

\REF\Strath{J. Strathdee, Int. Jour. Mod. Phys. {\bf A2} (1987) 273.}
\REF\GravDu{C.M. Hull, Nucl. Phys. {\bf B509} (1997) 252, hep-th/9705162.}
\REF\abo{M. Abou-Zeid,  B. de Wit,  D. Lust and  H. Nicolai,
Phys. Lett.  {\bf B466} (1999) 144, hep-th/9908169.}
\REF\roz{M. Rozali, Phys.Lett. {\bf  B400} (1997) 260, hep-th/9702136.}
\REF\brs{M. Berkooz, M. Rozali, and N. Seiberg, 
Phys.Lett. {\bf B408} (1997) 105,
hep-th/9704089.}
\REF\note{N.Seiberg, hep-th/9705117.}
 \REF\WitUSC{E. Witten, in Proceedings of Strings 95, hep-th/9507121.}
\REF\sei{N. Seiberg, 
Phys.Lett. {\bf B408} (1997) 98, hep-th/9705221.}
\REF\dvv{R. Dijkgraaf, E. Verlinde, and H. Verlinde,  hep-th/9603126;
 hep-th/9604055.} 
 \REF\lit{O. Aharony, hep-th/9911147.}
\REF\GHaw{G.W. Gibbons and S.W. Hawking, Phys. Lett. {\bf 78B} (1978) 430.}
\REF\GP{G.W. Gibbons and M.J. Perry, Nucl. Phys. {\bf B248} (1984) 629.}
\REF\SGP {R. Sorkin, Phys. Rev. Lett. {\bf 51} (1983) 87 ; D. Gross and
M. Perry, Nucl. Phys. {\bf B226} (1983) 29.}
\REF\CvH{ M. Cveti\v c and C.M. Hull, Nucl. Phys. {\bf B480} (1996) 296.}
\REF\CY{M. Cveti\v c and D. Youm, Nucl. Phys. {\bf B476} (1996) 118, hep-th/9603100.}

\def \aa {\alpha}
\def \bb {\beta}

\def \ll {\lambda}

\def \lll {\Lambda}

\def \www{\Omega}

\def \2 {{1 \over 2}}
\def \3 {{1 \over 3}}
\def \4 {{1 \over 4}}
\def \5 {{1 \over 5}}
\def \6 {{1 \over 6}}
\def \7 {{1 \over 7}}
\def \8 {{1 \over 8}}
\def \9 {{1 \over 9}}
\def \0 { \infty}

\def\ek {\eqn\abc$$}

\def \qq {\qquad}


%
%
%
%
\newhelp\stablestylehelp{You must choose a style between 0 and 3.}%
\newhelp\stablelinehelp{You should not use special hrules when stretching%
a table.}%
\newhelp\stablesmultiplehelp{You have tried to place an S-Table %
inside another%
S-Table.  I would recommend not going on.}%
%
%
\newdimen\stablesthinline
\stablesthinline=0.4pt
\newdimen\stablesthickline
\stablesthickline=1pt
%
%
\newif\ifstablesborderthin
\stablesborderthinfalse
\newif\ifstablesinternalthin
\stablesinternalthintrue
\newif\ifstablesomit
\newif\ifstablemode
\newif\ifstablesright
\stablesrightfalse
%
%
\newdimen\stablesbaselineskip
\newdimen\stableslineskip
\newdimen\stableslineskiplimit
%
%
\newcount\stablesmode
\newcount\stableslines
\newcount\stablestemp
\stablestemp=3
\newcount\stablescount
\stablescount=0
\newcount\stableslinet
\stableslinet=0
%
%
%
\newcount\stablestyle
\stablestyle=0
%
%
\def\stablesleft{\quad\hfil}%
\def\stablesright{\hfil\quad}%
%
%
\catcode`\|=\active%
%
%
\newcount\stablestrutsize
\newbox\stablestrutbox
\setbox\stablestrutbox=\hbox{\vrule height10pt depth5pt width0pt}
\def\stablestrut{\relax\ifmmode%
                         \copy\stablestrutbox%
                       \else%
                         \unhcopy\stablestrutbox%
                       \fi}%
%
%
\newdimen\stablesborderwidth
\newdimen\stablesinternalwidth
\newdimen\stablesdummy
\newcount\stablesdummyc
\newif\ifstablesin
\stablesinfalse
%
%
\def\begintable{\stablestart%
  \stablemodetrue%
  \stablesadj%
  \halign%
  \stablesdef}%
\def\stablesadj{%
  \ifcase\stablestyle%
    \hbox to \hsize\bgroup\hss\vbox\bgroup%
  \or%
    \hbox to \hsize\bgroup\vbox\bgroup%
  \or%
    \hbox to \hsize\bgroup\hss\vbox\bgroup%
  \or%
    \hbox\bgroup\vbox\bgroup%
  \else%
    \errhelp=\stablestylehelp%
    \errmessage{Invalid style selected, using default}%
    \hbox to \hsize\bgroup\hss\vbox\bgroup%
  \fi}%
\def\stablesend{\egroup%
  \ifcase\stablestyle%
    \hss\egroup%
  \or%
    \hss\egroup%
  \or%
    \egroup%
  \or%
    \egroup%
  \else%
    \hss\egroup%
  \fi}%
\def\stablestart{%
  \ifstablesin%
    \errhelp=\stablesmultiplehelp%
    \errmessage{An S-Table cannot be placed within an S-Table!}%
  \fi
  \global\stablesintrue%
  \global\advance\stablescount by 1%
  \message{<S-Tables Generating Table \number\stablescount}%
  \begingroup%
  \stablestrutsize=\ht\stablestrutbox%
  \advance\stablestrutsize by \dp\stablestrutbox%
  \ifstablesborderthin%
    \stablesborderwidth=\stablesthinline%
  \else%
    \stablesborderwidth=\stablesthickline%
  \fi%
  \ifstablesinternalthin%
    \stablesinternalwidth=\stablesthinline%
  \else%
    \stablesinternalwidth=\stablesthickline%
  \fi%
  \tabskip=0pt%
  \stablesbaselineskip=\baselineskip%
  \stableslineskip=\lineskip%
  \stableslineskiplimit=\lineskiplimit%
  \offinterlineskip%
  \def\borderrule{\vrule width \stablesborderwidth}%
  \def\internalrule{\vrule width \stablesinternalwidth}%
  \def\thinline{\noalign{\hrule height \stablesthinline}}%
  \def\thickline{\noalign{\hrule height \stablesthickline}}%
  \def\trule{\omit\leaders\hrule height \stablesthinline\hfill}%
  \def\ttrule{\omit\leaders\hrule height \stablesthickline\hfill}%
  \def\tttrule##1{\omit\leaders\hrule height ##1\hfill}%
  \def\stablesel{&\omit\global\stablesmode=0%
    \global\advance\stableslines by 1\borderrule\hfil\cr}%
  \def\el{\stablesel&}%
  \def\elt{\stablesel\thinline&}%
  \def\eltt{\stablesel\thickline&}%
  \def\elttt##1{\stablesel\noalign{\hrule height ##1}&}%
  \def\elspec{&\omit\hfil\borderrule\cr\omit\borderrule&%
              \ifstablemode%
              \else%
                \errhelp=\stablelinehelp%
                \errmessage{Special ruling will not display properly}%
              \fi}%
  \def\stmultispan##1{\mscount=##1 \loop\ifnum\mscount>3 \stspan\repeat}%
  \def\stspan{\span\omit \advance\mscount by -1}%
  \def\multicolumn##1{\omit\multiply\stablestemp by ##1%
     \stmultispan{\stablestemp}%
     \advance\stablesmode by ##1%
     \advance\stablesmode by -1%
     \stablestemp=3}%
  \def\multirow##1{\stablesdummyc=##1\parindent=0pt\setbox0\hbox\bgroup%
    \aftergroup\emultirow\let\temp=}
  \def\emultirow{\setbox1\vbox to\stablesdummyc\stablestrutsize%
    {\hsize\wd0\vfil\box0\vfil}%
    \ht1=\ht\stablestrutbox%
    \dp1=\dp\stablestrutbox%
    \box1}%
  \def\stpar##1{\vtop\bgroup\hsize ##1%
     \baselineskip=\stablesbaselineskip%
     \lineskip=\stableslineskip%

\lineskiplimit=\stableslineskiplimit\bgroup\aftergroup\estpar\let\temp=}%
  \def\estpar{\vskip 6pt\egroup}%
  \def\stparrow##1##2{\stablesdummy=##2%
     \setbox0=\vtop to ##1\stablestrutsize\bgroup%
     \hsize\stablesdummy%
     \baselineskip=\stablesbaselineskip%
     \lineskip=\stableslineskip%
     \lineskiplimit=\stableslineskiplimit%
     \bgroup\vfil\aftergroup\estparrow%
     \let\temp=}%
  \def\estparrow{\vfil\egroup%
     \ht0=\ht\stablestrutbox%
     \dp0=\dp\stablestrutbox%
     \wd0=\stablesdummy%
     \box0}%
  \def|{\global\advance\stablesmode by 1&&&}%
  \def\|{\global\advance\stablesmode by 1&\omit\vrule width 0pt%
         \hfil&&}%
  \def\vt{\global\advance\stablesmode by 1&\omit\vrule width
\stablesthinline%
          \hfil&&}%
  \def\vtt{\global\advance\stablesmode by 1&\omit\vrule width
\stablesthickline%
          \hfil&&}%
  \def\vttt##1{\global\advance\stablesmode by 1&\omit\vrule width ##1%
          \hfil&&}%
  \def\vtr{\global\advance\stablesmode by 1&\omit\hfil\vrule width%
           \stablesthinline&&}%
  \def\vttr{\global\advance\stablesmode by 1&\omit\hfil\vrule width%
            \stablesthickline&&}%
\def\vtttr##1{\global\advance\stablesmode by 1&\omit\hfil\vrule width ##1&&}%
\stableslines=0%
\stablesomitfalse}%
\def\stablesdef{\bgroup\stablestrut\borderrule##\tabskip=0pt plus 1fil%
  &\stablesleft##\stablesright%
  &##\ifstablesright\hfill\fi\internalrule\ifstablesright\else\hfill\fi%
  \tabskip 0pt&&##\hfil\tabskip=0pt plus 1fil%
  &\stablesleft##\stablesright%
  &##\ifstablesright\hfill\fi\internalrule\ifstablesright\else\hfill\fi%
  \tabskip=0pt\cr%
  \ifstablesborderthin%
    \thinline%
  \else%
    \thickline%
  \fi&%
}%
\def\endtable{\advance\stableslines by 1\advance\stablesmode by 1%
   \message{- Rows: \number\stableslines, Columns:
\number\stablesmode>}%
   \stablesel%
   \ifstablesborderthin%
     \thinline%
   \else%
     \thickline%
   \fi%
   \egroup\stablesend%
\endgroup%
\global\stablesinfalse}
%
%

 \def\unit{\hbox to 3.3pt{\hskip1.3pt \vrule height 7pt width .4pt \hskip.7pt
\vrule height 7.85pt width .4pt \kern-2.4pt
\hrulefill \kern-3pt
\raise 4pt\hbox{\char'40}}}

\def\nup#1({Nucl.\ Phys.\  {\bf B#1}\ (}


\Pubnum{ \vbox{ \hbox {QMW-00-02}  \hbox 
{RUNHETC-2000-11}\hbox{hep-th/0004086}} }
\pubtype{}
\date{April, 2000}

\titlepage

\title {\bf  BPS Supermultiplets in Five Dimensions}

\author{C.M. Hull}
\address{Physics Department,
Queen Mary and Westfield College,
\break
Mile End Road, London E1 4NS, U.K.}
\andaddress{Department of Physics and Astronomy, Rutgers University, Piscataway, New Jersey 
08855-0849,
USA}
\vskip 0.5cm

\abstract { BPS representations
 of 5-dimensional supersymmetry algebras are classified. 
For BPS states preserving 1/2 the supersymmetry, there are 
two distinct classes of multiplets for $N=4$ supersymmetry
and three classes for $N=8$ supersymmetry.
For $N=4$ matter theories, the two 1/2 supersymmetric
 BPS multiplets are the massive vector multiplet and the
massive self-dual 2-form multiplet.
Some applications to
super-Yang-Mills, supergravity and little string theories are
considered.
 }

\endpage

Representations of supersymmetry carrying central charges have been extensively studied; see 
e.g. [\Strath].
The purpose of this note is to clarify some of the issues that arise in 4+1 dimensions,
improving on some misleading statements in the literature,
 and to classify the possible BPS multiplets. 
Some standard results are recovered, but   some unexpected features
 are found, such as the existence of three distinct
possible
representations for massive BPS states in $N=8$ theories preserving 1/2 the supersymmetry.

The $N=2n$ extended superalgebra in 4+1 dimensions with scalar 
central charges
has automorphism group $Sp(n)=USp(2n)$ and the anti-commutator
$$
\{Q_\alpha^a,Q_\beta^b\} =\, \www^{ab}\big( \Gamma^MC\big)_{\alpha\beta} P_M 
 +
 C _{\alpha\beta}( Z^{ab}+\www^{ab}K)
\eqn\fialgo$$
Here
 $\aa=1,\dots ,4$ are spinor indices and
$a=1,\dots N$ are $Sp(n)$ indices, 
$C _{\alpha\beta}$ is the charge conjugation matrix and $\www^{ab}$ is the symplectic invariant 
of
$Sp(n)$.
The supercharges $Q_\alpha^a$ are
symplectic Majorana spinors 
satisfying
$$
(\bar Q)^\alpha_a= C^{\aa \bb} \www_{ab}
Q_\bb ^b
\eqn\real$$
 The central
charges
$Z^{ab}$ satisfy $Z^{ab}=-Z^{ba}$ and
$\www_{ab}Z^{ab}=0$.
This can be generalised by adding 1-form and 2-form central charges 
which are carried by
  1,2,3 and 4-branes [\GravDu].
In the following, it will be useful to define
  $$W^{ab}=Z^{ab}+\www^{ab} K
  \ek
  
   In super-Yang-Mills theories and in $N=8$ supergravity, the charges $Z^{ab}$
are electric charges while $K$ is 
carried by solitons in 4+1 dimensions that arise from lifting instantons 
in 4 Euclidean dimensions. In $N=4$ supergravity coupled to matter, $K$ is a linear combination 
of such an instantonic charge and an ordinary electric charge.
The massless representations have   vanishing central charges $W^{ab} =0$ and
are supermultiplets with $2^N$ states fitting into representations of
the little group $SO(3) \times Sp(n)$;  a list of supergravity and  supermatter
representations with $N \le 8$ is given in  [\Strath].
For example, the $N=4$ super-Yang-Mills multiplet
decomposes into the following representations of $SO(3) \times Sp(2)$:
$$
N=4 ~SYM: \qq 2^4=(3,1)+(1,5)+(2,4)
\eqn\sym$$
and there is a variant $N=4$ super-Yang-Mills multiplet, the $Sp(2)$ super-Yang-Mills multiplet,
with the following structure:
$$(3,10) +(1,5) +(1,10) + (1,35)+(2,4)+(2,16)+(2,20)
\ek
in  which the vector fields are in a {\bf 10} of the R-symmetry group $Sp(2)\sim Spin(5)$.
The $N=8$ supergravity multiplet 
decomposes into the      $SO(3) \times Sp(4)$ representations:
$$
N=8 ~SUGRA: \qq 2^8=(5,1)+(3,27)+(1,42)+(4,8)+(2,48)
\eqn\sugra$$

The general (non-BPS) massive representations 
have dimension $4^N$, but there are shorter representations in which the
central charge $W^{ab} \ne 0$   and the mass saturates a BPS bound.
The little group for the 
massive representations is
$Spin(4)\times Sp(n)$. 
The supercharges, which transform as a $(4,n)$ of $Spin(4,1)\times 
Sp(n)$ and satisfy
  the reality constraint \real,
transform as a $(2,1;n) + (1,2;n)$ under the little group
$SU(2)\times SU(2)\times Sp(n)$, giving
the \lq chiral' supercharges 
$Q_{\pm}^{a}$ with $Q_{\pm}^{a}= \pm\Gamma^0Q_{\pm}^{a}$.
In the rest frame with
$P^M=(M,0,...,0)$ (with signature $(+----)$), the algebra \fialgo\  takes the form
$$
\eqalign{&
\{Q_{+},Q_{+}^{\dagger} \} = M\unit _{2n\times 2n}-\hat W
\cr
&
\{Q_{-},Q_{-}^{\dagger} \} = M\unit _{2n\times 2n}+\hat W
\cr&
\{Q_{+},Q_{-}^{\dagger} \} =0
\cr}\eqn\fialgao$$
where
$\hat W = -\www ^{-1} W$.

It will be useful to choose a basis 
in which $W$ is skew-diagonal with skew eigenvalues 
$\ll_{1},\ldots,\ll_{n}$, so that
$$W=
\lll \otimes 
\pmatrix {0 & 1 \cr
-1 & 0 \cr}
\ek
where $\lll$ is the diagonal $n\times n$
matrix whose entries are the eigenvalues
$$\lll=diag (\ll_1,...\ll_n)
\ek
and the symplectic invariant is 
$$\www=\unit _{n\times n}\otimes 
\pmatrix {0 & 1 \cr
-1 & 0 \cr}
\ek
so that $\hat W$ is diagonal
 $$
\hat W
=\lll \otimes 
\pmatrix {1 & 0 \cr
0 & 1 \cr}
\ek
with eigenvalues $\ll_i$, each of degeneracy two.
Then the 
BPS bound derived from \fialgao\ is
$$M
\ge \ll , \qq \ll \equiv max \vert \ll _{i}\vert
\ek
Note that
$$
K=2 \sum _{i} \ll_{i}= tr \hat W
\ek
so that $\lll$ is traceless unless $K\ne 0$. If $Z\ne 0$, then the central charge $Z$ will break 
the $Sp(n)$ symmetry to a subgroup.

Let $r_{+}$ be the number of the eigenvalues $\ll_{i}$ satisfying
$\ll_{i}= \ll$ 
 and $r_{-}$ be the number of  $\ll_{i}$ satisfying
$\ll_{i}= -\ll$
(so that $ r_{+}+r_{-}\le n$). A BPS state satisfying $M=\ll$ will be 
invariant under the action of $2r_{+}$ of the 2-component supercharges
$Q_{+}$ and $2r_{-}$ of the 2-component supercharges
$Q_{-}$, i.e. it is invariant under $4r_{+}$ positive chirality 
supersymmetries and $4r_{-}$ negative chirality ones.
The fraction of supersymmetry preserved is
$\nu = (r_{+}+r_{-})/2n$. Note that  $\nu \le 1/2$.
The state will then fit into a representation 
of the supersymmetry algebra generated by
  the remaining $4p$ positive chirality 
supersymmetries and $4q$ negative chirality ones, with $p=n-r_{+}$, 
$q=n-r_{-}$; this will be referred to   as a $(p,q)$ massive $D=5$ supermultiplet. 

One motivation for this nomenclature is so that
the dimensional reduction on a circle  of a massless representation of  $(p,q)$  
supersymmetry in 5+1 dimensions gives a Kaluza-Klein tower of massive
multiplets in 4+1 dimensions, each of which is a $(p,q)$ massive $D=5$ supermultiplet.
(Recall that the $(p,q)$ superalgebra in 6 dimensions has
$p$ right-handed symplectic 
Majorana-Weyl 
supercharges and $q$ left-handed ones.)
The momentum in the circle direction gives the central charge carried 
by the massive multiplet on dimensional  reduction.
The   $(p,q)$ multiplet decomposes into representations of the
\lq little group'   $Spin(4)\times Sp(p)\times Sp(q)$.

There are similar multiplets in other odd dimensions; for example, in 9 dimensions, there
are (1,0), (2,0) and (1,1) massive BPS supersymmetry multiplets, such as those 
arising from the   Kaluza-Klein modes of circle compactifications of (1,0), (2,0) and (1,1) 
supersymmetric theories
 in $D=10$, respectively.
The BPS states of type II string theory compactified to 9 dimensions were considered in [\abo], where
three types of multiplets were identified, the KKA, the KKB and the intermediate multiplets; these are (1,1), (2,0) and (2,1) multiplets, respectively.

The massive multiplets with $N\le 8$ and $p\ge q$ are listed in table 1. 
There are two types of 1/2 supersymmetric  BPS multiplets 
with $N=4$, the (1,1) and (2,0) multiplets,  and three types with $N=8$,
the (4,0),(3,1) 
and (2,2) multiplets.

\vskip 0.5cm
{\vbox{
\begintable
   $N$| $r_+$ | $r_-$ | $(p,q)$ |$\nu$ \elt 
    $2n$  |  0      |     0 |     $(n,n)$    |  0     \elt
     2 |   0     |   1   |  (1,0)       |  1/2     \elt
     4 |   0     |   2   |   (2,0)      |    1/2   \elt
     4 |   1     |   1   |    (1,1)     |    1/2   \elt
     4 |  0      |   1   |   (2,1)      |     1/4  \elt
     6 |  0      |     1 |   (3,2)      |     1/6  \elt
   6   |  0      |   2   |   (3,1)      |    1/3   \elt
    6  |   1     |   1   |    (2,2)     |   1/3    \elt
    6  |  0      |  3    |     (3,0)    |    1/2   \elt
     6 | 1       |  2    |      (2,1)   |     1/2  \elt
    8  |  0      |  1    |    (4,3)     |   1/8    \elt
    8  |  0      | 2     |    (4,2)     |    1/4   \elt
    8  |   1     |   1   |  (3,3)       |      1/4 \elt
    8  |  0      |   3   |  (4,1)       |     3/8  \elt
    8  |      1  |  2    |   (3,2)      | 3/8      \elt
    8  |     0   |   4   | (4,0)        | 1/2      \elt
    8  |    1    |   3   | (3,1)        | 1/2      \elt
    8  |  2      | 2     |     (2,2)    | 1/2     
   \endtable

{\bf Table 1} {\it
BPS multiplets in 5 dimensions.}
The $(p,q)$ BPS multiplets for $N\le 8$ supersymmetry in 5 dimensions with $p\ge q$
are listed. They preserve a fraction $\nu$ of the $4N$
supersymmetries
and fit into  massive multiplets of the
active $(p,q)$ supersymmetry.
}}
\vskip .5cm

Consider first $N=4$ supersymmetry, with two eigenvalues $\ll_1,\ll_2$.
If $\ll_1 \ne \ll_2$, then the central charge $Z\ne 0$ and 
breaks the $Sp(2)\sim Spin(5)$ R-symmetry to $Sp(1)\times Sp(1)\sim Spin(4)$.
If for $i=1,2$,    $M > \vert \ll_i \vert$, no supersymmetry is preserved and the multiplet is 
the general (2,2)
massive one.
If $M=\vert \ll_1 \vert > \vert \ll_2 \vert$, then 1/4 supersymmetry is preserved
(i.e. 4 of the 16 supersymmetries)
and the multiplet is 
(2,1) or (1,2) depending on the sign of $\lambda _1$.
For 1/2 supersymmetric
states,
$M= \vert \ll_1 \vert=\vert \ll_2 \vert$.
The (1,1) BPS multiplets
have 
$\lll = diag( \ll , - \ll)$ and so $K=0$ but the 
 central charge $Z^{ab}$ is non-zero, breaking the
$SO(5)$ R-symmetry to $SO(4)$, while the
(2,0) multiplets have 
$\lll = - diag( \ll ,  \ll)$ and so 
$Z^{ab}=0$ and $K = -\ll$ is non-zero, preserving
the full $SO(5)$.
The   (1,1) supersymmetric massive vector multiplet has
the 
$SU(2)\times SU(2)\times Sp(1)\times Sp(1)$ content
$$(2,2;1,1)+(1,1;2,2)+(2,1;1,2)+(1,2;2,1)
\ek
while there is a (2,0) massive multiplet with
the 
$SU(2)\times SU(2)\times Sp(2)$ content
$$(3,1;1)+(1,1;5)+(2,1;4)
\ek

The massive covariant field whose physical degrees of freedom are in the 
 $(3,1;1)$ representation is 
a self-dual massive 2-form field, i.e. a 2-form gauge field
$B$ satisfying the self-duality constraint
$$
dB=m*B
\ek
where $*$ is the Hodge dual and $m$ is the mass.
Such massive self-dual $n$-form fields in $2n+1$ dimensions 
occur in various supergravity theories and their compactifications; for example, 
a massive self-dual 4-form occurs in the (2,0) massive multiplet in 9 dimensions and arises 
as a Kaluza-Klein mode for 
 the self-dual 4-form of $D=10$ IIB supergravity when compactified on a circle.
Note that    
these massive vector and massive 2-form
multiplets in $D=5$ both have the same massless limit, which is
the massless vector multiplet \sym\ (after dualising
the massless 2-form to a 1-form gauge field),
so that in $D=5$ there are two distinct massive generalisations
of the Yang-Mills multiplet.
There are also (1,1) and (2,0) massive supergravity and gravitino
multiplets that arise as Kaluza-Klein modes in the 
compactification  of 6-dimensional supergravity theories on a
circle.
 
The $N=4,D=5$ super-Yang-Mills theory with the multiplet \sym\
has  a vector field $A_M$  and 5 scalars $\phi ^{ab} =-\phi ^{ba}$, $\www_{ab}\phi^{ab}=0$ 
($a,b=1,\ldots ,4$), all in the adjoint of the gauge group.
The central charges in the algebra \fialgo\
are the
 five electric charges
$$ Z^{ab}= {1 \over 4\pi}\int tr (\phi ^{ab} *F)
\ek
integrated over a 3-sphere at spatial infinity
and the instanton charge
$$
K={4\pi ^2 \over g_{YM}^{2}}n_I, \qq n_I={1 \over 32\pi^2 }\int tr (F\wedge F)
\ek
integrated over a spatial hypersurface, where $n_I$ is the  integral instanton number  and    $g_{YM}$ is 
the Yang-Mills coupling constant.
The W-bosons carry electric charge and fit into (1,1) massive vector multiplets
while the 0-brane solitons in 4+1 dimensions arising
 from lifting Yang-Mills instantons in 4 Euclidean dimensions
carry the charge $K$ and fit into (2,0) massive
self-dual tensor multiplets.
There is a tower of instantonic 0-branes with  mass proportional to $\vert n_I\vert /g_{YM}^{2}$ 
for all integers $n_I$ which become light in the strong coupling limit 
$g_{YM}\to \infty$, 
 signalling that an extra dimension is opening up, with the tower of 
 massive states interpreted as a Kaluza-Klein tower for a (2,0) 
 supersymmetric theory in 6 dimensisions compactified on a circle [\roz-\note].
If a finite number of (1,1) massive multiplets become massless at
some point in moduli space, then  there is an
enhancement of the gauge symmetry
at that point. If an infinite number become massless, this can signal a 
decompactification
to a (1,1) supersymmetric theory in 6 
dimensions.
 
The  little string theory in 6 dimensions with (2,0)   supersymmetry [\brs-\lit]
 compactified
on a circle of radius $R$
has an infinite tower of momentum modes
which are in (2,0) D=5 massive multiplets
and an infinite tower of winding modes, which
are in (1,1) multiplets. T-duality takes this to the (1,1) little 
string theory on a circle with inverse radius $l_{s}^{2}/R$ for which
the (2,0) multiplets are now winding modes and the (1,1) ones are 
momentum modes.

The $N=8$ supergravity theory in $D=5$ dimensions has the multiplet 
structure \sugra\ including 27 abelian vector fields, and the electric 
charges for these (suitably dressed with scalars) give the 27 central charges $Z^{ab}$ in 
\fialgo.
The charge $K$ was defined in [\GravDu] and is carried by 1/2-supersymmetric solutions of the 
form $N\times \R$ where $N$ is a self-dual gravitational  instanton in 4 
Euclidean dimensions [\GHaw] and $\R$ is time [\GravDu,\GP]. If $N$ is the multi-Taub NUT 
instanton, the solution has an interpretation as a multi-Kaluza-Klein 
monopole space [\SGP], while if it is a multi-Eguchi Hansen space, the 
solution can be interpreted as representing a set of 0-branes   
in a 5-dimensional space-time [\GravDu].

It is straightforward to find the
decomposition of each multiplet into representations of the
little group $SU(2)\times SU(2)\times Sp(p)\times Sp(q)$, using
e.g. the methods of [\Strath].
For the three 1/2-supersymmetric multiplets that could arise in $N=8$
supergravity (i.e. which  reduce to $D=4$ multiplets with spins no greater than two),
the results
are as follows.
The (2,2) multiplet has the
$SU(2)\times SU(2)\times Sp(2)\times Sp(2)$ content
$$\eqalign{
&(3,3;1,1)+(1,3;5,1)+(3,1;1,5)+(1,1;5,5)+(2,2;4,4)
\cr &
+(2,3;4,1)+(3,2;1,4)+(2,1;4,5)+(1,2;5,4)
\cr}
\ek
with $$ \lll=diag(\ll,\ll,-\ll,-\ll)
\ek
so that 
$$K=0,\qq Z=
diag(\ll,\ll,-\ll,-\ll) \otimes 
\pmatrix {0 & 1 \cr
-1 & 0 \cr}
\ek
The (3,1) multiplet has the
$SU(2)\times SU(2)\times Sp(3)\times Sp(1)$ content
$$\eqalign{
&(4,2;1,1)+(3,1;6,2)+(1,1;14',2)+(2,2;14,1)
\cr &
+(3,2;6,1)+(2,1;14,2)+(1,2;14',1)
\cr}
\ek
with $$ \lll=diag(\ll,-\ll,-\ll,-\ll)
\ek
so that 
$$K=-4\ll ,\qq Z=
diag(3\ll,-\ll,-\ll,-\ll) \otimes 
\pmatrix {0 & 1 \cr
-1 & 0 \cr}
\ek
The (4,0) multiplet has the
$SU(2)\times SU(2)\times Sp(4)$ content
$$(5,1;1)+(3,1;27)+(1,1;42)+(4,1;8)+(2,1;48)
\ek
with $$ \lll=-diag(\ll,\ll,\ll,\ll)
\ek
so that $Z=0$ and $K=-8\ll$.

If $K=0$, the only possibilities for BPS states are those  preserving 1/2 the 
supersymmetry and fitting into massive (2,2) multiplets, the 1/4 
supersymmetric states in (3,3) supermultiplets and the 1/8 supersymmetric 
states in (4,3) or (3,4) multiplets; all the other possibilities for 
$N=8$ that are listed in table 1, including 3/8 supersymmetry, cannot occur 
with $K=0$. The general BPS 0-brane or black hole solutions 
with $K=0$ were given in [\CvH] by acting on the generating solution of [\CY] 
with U-duality transformations.

A non-zero value for $Z^{ab}$ necessarily breaks the $Sp(4)$ to
a subgroup.
The instantonic solutions with $Z^{ab}=0$ and $K\ne 0$
fit into (4,0) or (0,4) multiplets, and the full $Sp(4)$ is preserved.
It is intriguing that the algebra also allows BPS states with both
$K$ and $Z$ non-zero, and which would fit into multiplets with
(4,2),(4,1),(3,2) or (3,1) supersymmetry. Of particular note is the unexpected 1/2 supersymmetric (3,1) multiplet,
as 1/2 supersymmetric states 
seem to play a special role in M-theory. 
The physics associated with these three 1/2-supersymmetric multiplets will be discussed 
in a separate publication.

\refout

\bye